\documentclass[acmsmall]{acmart}

\usepackage{xcolor}
\definecolor{blue}{rgb}{0, 0, 0}

\AtBeginDocument{%
  \providecommand\BibTeX{{%
    \normalfont B\kern-0.5em{\scshape i\kern-0.25em b}\kern-0.8em\TeX}}}

\begin{document}

\title{Combating Misinformation in Bangladesh: Roles and Responsibilities as Perceived by Journalists, Fact-checkers, and Users}

\author{Md Mahfuzul Haque}
\affiliation{%
 \streetaddress{Philip Merrill College of Journalism}
 \institution{University of Maryland}
 \city{College Park}
 \state{Maryland}
 \country{USA}}

\author{Mohammad Yousuf}
\affiliation{%
 \streetaddress{Department of Communication and Journalism}
 \institution{University of New Mexico}
 \city{Albuquerque}
 \state{New Mexico}
 \country{USA}}

\author{Ahmed Shatil Alam}
\affiliation{%
 \streetaddress{School of Journalism and New Media}
 \institution{University of Mississippi}
 \city{Oxford}
 \state{Mississippi}
 \country{USA}}

\author{Pratyasha Saha}
\affiliation{%
 \streetaddress{Department of Physics}
 \institution{University of Dhaka}
 \city{Dhaka}
 \country{Bangladesh}}

\author{Syed Ishtiaque Ahmed}
\affiliation{%
 \streetaddress{Department of Computer Science}
 \institution{University of Toronto}
 \city{Toronto}
 \state{Ontario}
 \country{Canada}}

\author{Naeemul Hassan}
\affiliation{%
 \streetaddress{Philip Merrill College of Journalism, College of Information Studies}
 \institution{University of Maryland}
 \city{College Park}
 \state{Maryland}
 \country{USA}}

\renewcommand{\shorttitle}{Combating Online Misinformation in Bangladesh}
\renewcommand{\shortauthors}{Haque et al.}

\begin{abstract}
\textcolor{blue}{There has been a growing interest within CSCW community in understanding the characteristics of misinformation propagated through computational media, and the devising techniques to address the associated challenges. However, most work in this area has been concentrated on the cases in the western world leaving a major portion of this problem unaddressed that is situated in the Global South. This paper aims to broaden the scope of this discourse by focusing on this problem in the context of Bangladesh, a country in the Global South.} The spread of misinformation on Facebook in Bangladesh, a country with a population of over 163 million, has resulted in chaos, hate attacks, and killings. By interviewing journalists, fact-checkers, in addition to surveying the general public, we analyzed the current state of verifying misinformation in Bangladesh. Our findings show that most people in the `news audience' want the news media to verify the authenticity of online information that they see online. However, the newspaper journalists say that fact-checking online information is not a part of their job, and it is also beyond their capacity given the amount of information being published online every day. We further find that the voluntary fact-checkers in Bangladesh are not equipped with sufficient infrastructural support to fill in this gap. \textcolor{blue}{We show how our findings are connected to some of the core concerns of CSCW community around social media, collaboration, infrastructural politics, and information inequality.} From our analysis, we also suggest several pathways to increase the impact of fact-checking efforts through collaboration, technology design, and infrastructure development. 
\end{abstract}

\ccsdesc[500]{Human-centered computing~Empirical studies in collaborative and social computing}
\keywords{Misinformation; Disinformation; Information Disorder; Bangladesh; ICTD}

\setcopyright{acmlicensed}
\acmJournal{PACMHCI}
\acmYear{2020} \acmVolume{4} \acmNumber{CSCW2} \acmArticle{130} \acmMonth{10} \acmPrice{15.00}\acmDOI{10.1145/3415201}
\maketitle

\section{Introduction}
\label{introduction}
\textcolor{blue}{Online misinformation is considered a major challenge for democracies as exposure to it tends to mislead people. This interferes with making informed decisions \cite{humprecht2020resilience}. Although misinformation is a global crisis, measures to combat it vary with respect to geographical location, culture, socioeconomic context, and political situation. CSCW (Computer-Supported Collaborative Work and Social Computing) researchers examined the spread of misinformation and offered possible solutions, mostly in the context of the Global North \footnote{Global North refers to the countries in North America and Western Europe, and also Japan and Australia. These rich countries are generally known as the `First World' \cite{muller2020search}. } (e.g., \cite{starbird2019disinformation,flintham2018falling,arif2017closer,jiang2018linguistic}). There is a gap in the literature on misinformation in the Global South \footnote{The term `Global South' generally refers to low or middle income countries that are located in Africa, Asia, Oceania, Latin America, and the Caribbean. These countries have a history of colonialism and are newly industrialized or in the process of industrialization \cite{dados2012global}.}. To this end, this research focuses on Bangladesh to understand people's perception of misinformation, how it is being combated, and what else can be done to fight against misinformation. By focusing on Bangladesh, this research enriches misinformation scholarship by adding knowledge from a country, where online misinformation has become a growing concern.}

Bangladesh, a South Asian country with a population of over 163 million, has experienced several incidents of mob lynching and hate attacks caused by the spread of inaccurate and misleading information on Facebook. For example, a series of mob attacks killed eight people in July 2019 after false rumors about child abduction spread online \cite{BBC2019}. The victims were targeted over rumors that human sacrifices were needed to build the Padma Bridge, a \$3 billion project to connect the southern region of Bangladesh to its capital Dhaka. The rumor was that a group of child abductors were kidnapping children across the country to sacrifice them to build the bridge. The victims of the mob attacks were suspected by ordinary citizens of having been on the lookout for abducting children~\cite{BBC2019}. Another instance of misinformation leading to violence took place in 2012, where a Facebook post of a local Buddhist youth included a photo of a burned copy of the Quran, the Muslim holy book, that triggered an angry mob to set fire to Buddhist temples, loot and vandalize more than a hundred homes in a Buddhist village at Ramu in Cox's Bazar \cite {Alam2012ramu}. It was later found that the image was photoshopped and someone else posted the image on the youth's Facebook page. Other Facebook users mistakenly thought that the youth posted the image to defame Islam \cite {Siddique2013arson}. Fake news has also been a growing concern in Bangladesh where duplicate websites mimicking some legacy news organizations were created to spread misleading political news during the country's parliamentary elections in 2018 \cite {Alam2018fakenews}. During that period, Facebook and Twitter also suspended several pages and accounts to curtail the spread of misleading news \cite {Siddiqui2018facebook}. Despite having such serious and widespread impact, the ecology of misinformation in Bangladesh has not been studied thoroughly.

The problem of rumors and misleading information is not unique to Bangladesh; other developing countries have experienced damaging consequences. For example, a doctored video of child kidnapping in India went viral on WhatsApp which created confusion among the general public. The video triggered mob attacks to kill at least two dozen people across India in 2019 \cite{Bengali2019}. Other developing countries such as Indonesia, Sri Lanka, the Philippines, Taiwan, Singapore, Vietnam have also suffered from misinformation \cite{kaur2018information,Harris2018srilanka}. \textcolor{blue}{Although the consequences have similarities across countries, resilience and preparation to combat misinformation differ from a country to another \cite{humprecht2020resilience}. So knowledge about misinformation from local contexts would benefit CSCW researchers. In this research, we interviewed journalists and fact checkers, in addition to conducting a survey with the general public in Bangladesh to understand how misinformation is being combated and to suggest possible solutions. Thus, this research offers the following contributions to CSCW literature:}

\textcolor{blue}{\begin{itemize}
    \item This work provides a thorough understanding of the spread of online misinformation in Bangladesh and associated complexities. The findings of this paper thus add knowledge to the CSCW scholarship around the global perspective on the misuse of information and communication platforms.
    \item This study offers lessons on the strategies for combating misinformation and associated challenges. This study shows how these challenges are connected to a wider set of political issues that often remain invisible to us. 
    \item Finally, this study offers recommendations about design of technology to advance fact-checking practice in a resource-constrained context, like Bangladesh.
\end{itemize}}


\section{Related Work}
\label{literature_review}

\subsection{Misinformation: A Global Challenge}
\label{misinfoaglobalchallenge}
The problem of misinformation has recently been an important topic for CSCW scholarship as this problem has become a pressing issue across the world \cite{humprecht2020resilience}. The problem, however, is not new. Misinformation, rumors, and propaganda are attributes of human communication that can be traced back to the Roman times when Antony and Cleopatra met. A misinformation campaign was designed to besmirch Antony~\cite{posetti2018short}. In 1493, when the Gutenberg printing press was invented, the spread of misinformation amplified significantly; one earlier example of large scale hoax is \textit{The New York Sun}'s `Great Moon Hoax' in 1835~\cite{vida2012great}. As \textit{The Guardian} columnist Natalie Nougayr\`ede has observed: ``The use of propaganda is ancient, but never before has there been the technology to so effectively disseminate it'' \cite{nougayrede2018age}. According to World Economic Forum~\cite{wef2016} and Edelman Trust index~\cite{eti}, the spread of misinformation through social networks is one of the key problems gripping the world. 

\textcolor{blue}{Misinformation efforts have targeted elections and policy issues of the countries in the Global North. Russian interference in the 2016 U.S. presidential elections generated widespread concerns. It is believed that the Russian government made concerted effort to meddle with the elections in the forms of online political propaganda, targeted advertisements to U.S. voters via social media \cite{badawy2018analyzing}. Russia is also believed to have conducted strategic information operations against the U.S. targeting political discourses, such as the \#BlackLivesMatter movement \cite{starbird2019disinformation}. Starbird et al. \cite{starbird2019disinformation} found that the Russian Internet Research Agency (IRA) targeted, infiltrated, and cultivated politically active communities on Twitter to interfere with the movement. During the 2018 U.S. midterm elections, misinformation efforts targeted Latino voters \cite{flores20192018}. Analyzing Reddit posts, Flores-Saviaga and Savage \cite{flores20192018} observed that those posts lacked neutral actors to engage Latinos in political topics. The posts included the rhetoric that President Trump is either anti or pro-Latinos. During the UK referendum on EU membership, known as the Brexit, the spread of misinformation caused much concerns. Political bots on Twitter were designed to strategically target the Brexit conversations \cite{howard2016bots}. Humprecht et al. \cite{humprecht2020resilience} found differences in terms of resilience and preparation to combat misinformation between North American and European countries. European countries are more resilient to combat misinformation. On the other hand, the situation in the U.S. is worse than the European countries due to political polarization and less trust in media \cite{humprecht2020resilience}.} 

\textcolor{blue}{Latin American countries also suffered from the spread of misinformation. For example, during the 2018 presidential elections in Brazil, false information favoring Jair Bolsonaro went viral on WhatsApp \cite{avelar2019whatsapp}. African countries also experienced damaging consequences from the spread of misinformation. For example, in South Africa, there were rumors that the coronavirus only affected white people. This led some residents to ignore warnings of getting infected. The South African government has criminalized the spread of misinformation about the virus \cite{wild2020africa}.}

\textcolor{blue}{In neighboring countries of Bangladesh, such as India, Myanmar, and Sri Lanka, misinformation efforts targeted religious sentiments~\cite{hussain2020infrastructuring, pal2016twitter, rajadesingan2020leader}. In India where Hindus are the majority, misinformation targeted the minority Muslims. Fabricated images and news of cow slaughtering by Muslim men went viral on WhatsApp which resulted in killing of Muslim men \cite{kaur2018information}. Similarly, in Sri Lanka and Myanmar, where Buddhists are the majority, Muslims were targeted during the 2019 bombing in Sri Lanka and Rohingya crisis in Myanmar \cite{kaur2018information}. In sum, the literature review suggests some patterns across countries in the developed and developing countries. In North American and European countries, misinformation efforts mostly targeted elections and political discourses. The same pattern also exists in Latin American and African countries. However, in South Asian countries, such as Bangladesh, India, Myanmar, and Sri Lanka, misinformation efforts mostly targeted religious sentiments. In response to this pressing problem, countries have taken various measures to combat misinformation.}


\subsection{Combating Misinformation: Approaches}
Governments, journalists, fact-checkers, educators and other entities across the world have taken various steps and are following certain strategies to combat misinformation. Below, we present various approaches currently employed to combat the problem.

\subsubsection{Fact-checking}
The practice of fact-checking began in the United States newspapers to verify claims of politicians as early as in 1980s \cite{graves2018boundaries}. The campaign of Ronald Reagan in 1979-1980 had been an impetus to the rise of fact-checking \cite{dobbs2012rise}. During the campaign for the presidential race, Reagan claimed that trees caused four times more pollution than automobiles. When Reagan became the president, reporters began verifying his press conferences and television statements, often adding a sidebar to the coverage of him, although the practice did not sustain \cite{dobbs2012rise}. Fact-checking was also present in the 1990s when ``Adwatch'' reports verified claims made by political advertisements \cite{frantzich2002watching}. Later, fact-checking began to develop when the \textit{FactCheck.org} was founded in 2003 with a pool of staff who were professional journalists. \textit{PolitiFact} and the Washington Post's \textit{Fact Checker} were established in 2007. \textit{PolitiFact} and \textit{Factcheck.org} played a key role in popularizing fact-checking when they got Pulitzer prize and some leading newspapers quoted their fact-checks. The organizations collaborated with newspapers so as to introduce fact-check in journalists' regular work \cite{amazeen2013making}. The movement became transnational in the subsequent years. Fact-checkers around the world meet annually at the conference of the IFCN (International Fact-checking Network), a platform of the Poynter Institute, where they share their experience and challenges \cite{graves2018boundaries}. Currently, 188 fact-checking organizations in 60 countries have been working as per the Duke Reporters' Lab \cite{stencel2019number}. However, due to the lack of sustainable business model, many fact-checking organizations cannot sustain for long~\cite{hassan2019examining}.

\subsubsection{Increasing Media Literacy} Making people media literate to combat the spread of misinformation is considered a way in which people are provided with a set of critical thinking skills to evaluate media content \cite{bulger2018promises}. Media literacy refers to a set of knowledge, skills, and habits of mind required for full participation in a contemporary media-saturated society \cite{hobbs2019media}. Although misinformation has affected people across the world, the initiatives on media literacy mostly confined to the developed nations. For example, in the United States, most of the media literacy efforts focus on teachers training and curricular development. The teacher training programs relating to media literacy include the Media Education Lab at the University of Rhode Island and the Project Look Sharp at Ithaca \cite{bulger2018promises}. Some news organizations such as \textit{The New York Times} and \textit{Washington Post} offer curricular resources around information credibility, use of evidence, and news production \cite{bulger2018promises}. As a way to combat misinformation, other developed nations such as Germany and Australia have introduced media literacy education at the school level \cite{tulodziecki2019media,hipfl2019media}. 

It should be noted here that media literacy is also dependent on tech-literacy and literacy in general. Designing appropriate technologies for low-literate people still remains a challenge for researchers in HCI4D and ICTD areas ~\cite{medhi2006text, medhi2011designing, ahmed2013ecologies, ahmed2015suhrid}. Despite these challenges, a few steps have been taken to make people aware about misinformation. In December 2018, ahead of the general elections, WhatsApp ran a television campaign in India to fight against circulation of fake news and asked users to check the integrity of the information they receive \cite{bali2019fake}. In another initiative, local police of Mahbubnagar district in the Telangana state in Southern India used town criers and staged folk songs to fight against misinformation and stop hate attacks caused by it \cite{Joglekar2018folk}. Although Bangladesh has similarly been suffering from the spread of misinformation, to the best of the authors' knowledge no media literacy program is in effect to combat misinformation. 

\subsubsection{Policy and Regulation} As people across the world have suffered from misinformation, many countries have taken measures to combat it. For example, the counter-propaganda agency named the EU East StratCom Task Force was created in 2015 at the direction of the European Council to respond to Russia's disinformation campaigns. The agency debunked thousands of cases since it was established \cite{stray2019institutional}. The Chinese government also took a policy of engaging the general public to combat misinformation. A 2014 report stated that two million people were involved in online content monitoring \cite{stray2019institutional}. Some European countries such as Italy have criminalized sharing and posting of `false, exaggerated or biased' information and imposed fines of up to 5,000 Euros \cite{bali2019fake}. Many other countries across North America, Europe, Asia, and Africa have also enacted laws and taken policy measures to combat misinformation \cite{funke2019misinformation}.     

\subsubsection{Computational Approaches}
Researchers across different disciplines have proposed computational, algorithmic approaches to combat misinformation. For instance, a group of researchers, primarily from the University of Texas at Arlington, developed \textit{ClaimBuster}, an end-to-end fact-checking system, that can automatically detect check-worthy factual claims and check with respect to already fact-checked claims or knowledge bases~\cite{hassan2017claimbuster}. They leveraged machine learning, natural language processing, and database query techniques in designing the system. \textit{ClaimBuster} is constantly monitoring factual claims from various sources including interview transcripts, social media, congressional records, etc. These claims are then sent to a political fact-checking organization, \textit{PolitiFact.com} through email. The use of \textit{ClaimBuster} demonstrated that automated reporting tools can handle important journalism tasks that reduce editorial workloads. However, a human touch is needed to increase the effectiveness of algorithms that alert editors of possible political falsehood~\cite{adair2019human}.

Researchers from Indiana University has developed several tools to detect misinformation spread by social bots~\cite{yang2019arming,lou2019information}. For instance, \textit{Hoaxy}~\cite{shao2016hoaxy} can monitor social networks and visualize how a fake claim and its related fact-checks spread on Twitter. The \textit{Botometer} (formerly BotOrNot)~\cite{davis2016botornot} can detect whether a Twitter account is a bot or not. These tools use natural language processing and artificial intelligence to find patterns in text data. Some researchers proposed knowledge-based solutions. For example, Shu et al.~\cite{shu2017fake,shu2018fakenewsnet} analyzed the misinformation problem from data mining perspective and developed a benchmark dataset for misinformation detection. Others leveraged existing crowd based knowledge networks such as Wikipedia to detect misinformation~\cite{ciampaglia2015computational, shi2016discriminative}. While the computational techniques are excellent as starting steps towards more robust misinformation combating systems, they need some facilities such as access to information, smart data infrastructures, freedom of press and so on, which might not be available in certain regions, particularly in many developing countries. There exists a knowledge gap regarding how computational solutions for combating misinformation should be designed where such facilities do not exist.

\textcolor{blue}{Researchers have examined the affordances of social networking sites to detect misinformation. Metaxas et al. \cite{metaxas2015using} examined the application of \textit{Twittertrails}, an interactive, web-based tool (twittertrails.com) that allows users to investigate the origin and propagation characteristics of a rumor and its refutation on Twitter. The findings suggest that the tool can collect relevant tweets and detect its originator, when and how the story broke, and propagators. CSCW researchers also examined the roles crowd can play to detect and debunk misinformation \cite{arif2017closer,jiang2018linguistic,ghenai2017catching,kriplean2014integrating,roitero2018many}. These and many other such works demonstrating the research on misinformation have been an active interest within CSCW community, but most of these studies lack the perspective of the Global South. Hence, many important infrastructural challenges and regional politics have been missing in this vein of CSCW work.}

\subsection{Journalistic Values and Fact-checking}
Academics and professional organizations proposed long lists of objective and subjective ethical criteria that constitute excellence in journalism \cite{bogart2004reflections,burgoon1982world,merrill1968elite,craig2010excellence}. To mention a few, the subjective criteria include integrity, good writing, and professionalism while objective criteria include independence, and public welfare \cite{merrill1968elite}. Bogart \cite{bogart2004reflections} offered a combined list of objective and subjective criteria that includes ``integrity, fairness, balance, accuracy, comprehensiveness, diligence in discovery, authority, breadth of coverage, variety of content, reflection of the entire home community, vivid writing, attractive makeup, packaging or appearance, and easy navigability.'' The Society of Professional Journalists (SPJ) listed four principles in its ethics code as ``the foundation of ethical journalism'': seeking truth, minimizing harm, acting independently, and being accountable and transparent \cite{spj2014code}.

\textcolor{blue}{Kovach and Rosenstiel identified truth and independence as journalism’s first obligations and verification as its essence \cite{kovach2007rosenstiel}. They noted that journalism seeks a “practical and functional form of truth” as “all truths -- even the laws of science -- are subject to revision”. The journalistic process of seeking truth begins with gathering and verifying facts and then providing a logical conclusion of their meaning. Though some principles such as fairness, balance, and objectivity are considered universal journalistic values, they are not prominent in either the SPJ ethics code or the Kovach and Rosenstiel’s book, two key references for journalistic values. Kovach and Rosenstiel described fairness and balance as ideas “too vague to rise to the level of essential elements” of journalism \cite{kovach2007rosenstiel}. Objectivity is another such an idea that is often used by critics to make a case against fact-checking. As Kovach and Rosenstiel noted, “The concept of objectivity has been so mangled it now is usually used to describe the very problem it was conceived to correct” p. 6 \cite{kovach2007rosenstiel}.}

\textcolor{blue}{The ambiguity regarding objectivity led to disagreements in the journalism community over fact-checking. Many news outlets across the world embraced fact-checking while scholars described it as a “genre of journalism” that serves some core journalistic duties such as monitoring power and providing accurate information \cite{brandtzaeg2018journalists, coddington2014etal, hofseth2016make, graves2016deciding}. Meanwhile, critics argue that fact-checking violates the traditional objectivity principle by adjudicating claims. Refraining from adjudicating claims had been part of news organizations’ routines for decades, which were aimed at protecting journalists and news organizations from “occupational hazards such as libel suits” \cite{shoemaker&reese1996}. For example, the press carried U.S. Senator McCarthy’s false accusations against communists in the 1950s, although reporters knew those accusations were false. It was done to maintain “the resilient objective routine” and keep reporting patterns intact \cite{shoemaker&reese1996}. The most sustained criticism of fact-checking is that it ignores the value-laden nature of political debates and seeks to recast discourses as lies and facts instead of differences in worldviews \cite{graves2017epistemology}. However, Kovach and Rosenstiel’s definition of truth as practical and function debunks this claim \cite{kovach2007rosenstiel}. Graves added that “In newswork routines, internal discourse, and public statements, fact checkers demonstrate constantly that they appreciate the ambiguous character of facts in politics” \cite{graves2017epistemology}.}

\textcolor{blue}{Though fact-checking is rooted in journalism and many fact-checkers working in independent organizations are celebrated award-winning journalists, fact-checkers perceive their roles and contributions to society to be unique and different from the roles and contributions of journalists \cite{graves2016journalists, shinthorson2017}. Having a network (the International Fact-Checking Network at Poynter) and an ethics code of their own supports this argument. Lowrey \cite{lowrey2017} found that fact-checking organizations were developing sustainable new practices and showing signs of being institutionalized. Further analysis of the differences between journalism and fact-checking is beyond the scope of this study.}

\section{ICT, Journalism, and Freedom of the Press in Bangladesh}
\label{background}

\subsection{Connectivity}
Bangladesh got connected to the Internet in 1996, offering access to online content and e-mail communication \cite{azad1997overview}. However, Internet use was confined to urban areas due to its limited availability and higher price of internet data \cite{hasnayen2016internet}, and is often cut off by the government~\cite{bin2017internet}. The Internet use experienced a dramatic growth during the 2010s when mobile phones and internet data became affordable. As of August 2018, the total internet connections reached 90 million, among which around 85 million accessed the Internet through mobile phones \cite{star2018active}. The government of Bangladesh adopted the Information, Communication and Technology (ICT) policy in 2009 by coining the slogan ``Digital Bangladesh'' to bring about socioeconomic transformation through ICT. As part of its effort, the government has digitized some of its public services such as online submission of tax returns, online tender, registration for overseas jobs \cite{rahman1digital}. Bangladesh earned nearly \$800 million in 2017, exporting locally made software and providing services like outsourcing and freelance work \cite{star2017ict}. \textcolor{blue}{This connectivity has turned out to be a double-edged sword because the Internet has contributed to human and economic development, but at the same time, this has also often plunged people of Bangladesh into some other threats (see ~\cite{ahmed2017privacy, mizan2019silencing, rahman2019art, nova2019online, hassan2019nonparticipation, moitra2020understanding}, for example), including this important one: exposure to online misinformation.}      

\subsection{ICT Policies and Politics}
\label{ict}
\textcolor{blue}{Article 39 of the Constitution of Bangladesh guaranteed right to expression and freedom of the press; however, governments have enacted some laws that are conflicting with Article 39 \cite{ahmed2009}. To regulate digital communications, the government of Prime Minister Khaleda Zia enacted the Information and Communication Technology (ICT) Act in November 2006. Section 57 of the Act caused concerns about its ostensible goal to restrict freedom of expression \cite{house2017freedom}. The Section authorized prosecution of anyone who publishes, in electronic form, material deemed fake, obscene, defamatory, or any material that tends to deprave or corrupt its audience \cite{reuters2018bd}. In 2013, Sheikh Hasina’s government toughened the ICT Act, eliminating the need for arrest warrants \cite{reuters2018bd}. Between June 2016 and May 2017, 300 people were arrested under the ICT Act and 19 journalists were implicated. The government also blocked several messaging apps and blogs to curb the spread of misinformation. In August 2016, 35 news websites were blocked for publishing objectionable contents about the government \cite{house2017freedom}. The government of Prime Minister Sheikh Hasina nullified the 2006 ICT Act and passed a new law called the Digital Security Act in 2018 \cite{digital2018act}. This Act also has sections that impose restrictions on freedom of expression. According to news reports and human rights organizations, this Act is used to suppress freedom of expression \cite{bangladesh2020laws,amnesty2019bd}. The government charged or arrested 20 journalists on charge of violating this law in less than five weeks in April and May 2020 \cite{bangladesh2020laws}. According to the Reporters Without Borders, the government “has a custom-made judicial weapon for silencing troublesome journalists – the 2018 digital security law, under which `negative propaganda' is punishable by up to 14 years in prison. As a result, self-censorship has reached unprecedented levels because editors are reluctant to risk imprisonment or their media outlet’s closure” \cite{reporters2020borders}. Although this law was enacted to control the spread of online misinformation, the government has been criticized for using the law to control dissent voices \cite{AmnestyInternational,Mahmud2018bangladesh,Woollacott2018bangladesh,mamun2018act}.}

\textcolor{blue}{Governments have also used various laws to control the press \cite{ahmed2009}. The level of freedom Bangladeshi media enjoyed fluctuated over time. Media was under tight government control in the first two decades since Bangladesh had gotten independence from Pakistan in 1971. The media started enjoying some freedom and flourishing after a revolution that ousted the country’s then autocratic ruler and installed a democratically-elected government in 1991 \cite{anam2002}. Mahfuz Anam, editor of \emph{The Daily Star}, an English-language daily in Bangladesh, had once suggested that the best gift for Bangladesh’s democracy was press freedom\cite{anam2002}. However, press freedom in the country kept worsening in the last decade. Organizations tracking press freedom across the world suggest the situation in Bangladesh has continued to worsen since 2009. On the World Press Freedom index, Bangladesh slid to 151$^{st}$ position in 2020 -- worse than Afghanistan, Pakistan, Russia, Venezuela -- from 121$^{st}$ position in 2009 \cite{rsf2009index,rsf2020ranking}.
    }
\section{Methodology}
\label{methods}

This research aims to understand the roles and responsibilities perceived by the masses, journalists, and fact-checkers to combat misinformation in Bangladesh. \textcolor{blue}{We applied a multi-stakeholder approach in this research. There are both quantitative and qualitative research elements in this study. We gathered qualitative data from semi-structured interviews with Bangladeshi journalists, fact-checkers, and users. The quantitative data were derived from a survey with the internet users.} \textcolor{blue}{One rationale for using the mixed methods approach is that while the survey enabled us to reach a large number (N=500) of internet users, a closer look into the demographics revealed that most of the users were 
educated (78.8\% having at least a bachelors degree). So, we conducted in-person interviews to gain the perspectives of other demographics such as house maids, drivers who traditionally don't have higher education and are also hard to reach through online-based surveys. Another reason is, an in-person interview enabled us to ask open-ended questions which gave us detailed views of their exposure to online misinformation.} Below, we describe the questionnaire development, participant recruitment, interview and survey protocol, and data analysis methods in detail.

\textcolor{blue}{\subsection{Questionnaire Development}}

\textcolor{blue}{We developed four separate sets of questionnaires to conduct the survey and interviews. While developing the questionnaires, we have taken into consideration the literature related to journalism practices~\cite{graves2018boundaries, amazeen2020journalistic}, fact-checking mission and process~\cite{graves2017epistemology, walter2019fact}, perceived credibility of news~\cite{fogg2003prominence, metzger2013credibility} and crowd-source fact-checking~\cite{liaw2013maater, nguyen2018believe, hassan2019examining}. Four authors of this paper collaboratively developed the questionnaires. First, the questionnaire for fact-checkers was designed to understand their work, verification process, and challenges. To develop the questionnaire, we reviewed literature relating to fact-checkers' work in the context of North America, Europe, Latin America, Africa, and Asia (e.g., \cite{graves2016understanding,graves2018boundaries}). We piloted the questionnaire with two fact-checkers in Bangladesh. Based on their feedback, we revised some questions and then finalized the questionnaire. Second, to develop the questionnaire for journalists, we reviewed related literature about fact-checking practice in journalism (e.g., \cite{graves2016deciding,lowrey2017}). We piloted the questionnaire with two journalists to finalize it. Third, the survey questionnaire that included a mix of multiple choices, Likert items~\cite{joshi2015likert}, and some open-ended questions was designed to gather data about how internet users in Bangladesh confront misinformation. The survey was reviewed by survey experts so as to ensure the best practices were followed for obtaining valid, high-quality data~\cite{redmiles2017summary}. After piloting the survey, we modified the language of some questions and answer choices. Fourth, we developed another questionnaire for semi-structured interviews with users. The survey questionnaire guided us to develop the interview questionnaire that allowed interviewees to offer detailed viewpoints about their exposure to misinformation. The Institutional Review Board (IRB) of the authors' institutions examined and approved the questionnaires. In the appendix section, we have included the interview protocols. The survey questionnaire can be found to this link \footnote{forms.gle/mieTv7gy4fQrpX2u9}.}

\subsection{Participant Recruitment}

\subsubsection{Fact-checkers}
There are three active fact-checking organizations in Bangladesh- BDFactCheck (www.bdfactcheck.com), Jaachai (www.jaachai.com), and Fact Watch (www.fact-watch.org). \textcolor{blue}{Both BDFactCheck and Jaachai have three fact-checkers respectively. Fact Watch has nine fact-checkers. These organizations have fact-checked 542, 110, and 60 claims between 2017 and 2019 respectively. We reached out to these organizations through their Facebook Messenger account for interviews. We also contacted them via email. Seven of them responded and agreed to give interviews.} Three fact-checkers were from BDFactCheck, three from Fact Watch, and one from Jaachai. All the interviewees were male.  

\subsubsection{Journalists}
There are print, broadcast, and online-only news organizations in Bangladesh. The print and broadcast media also offer their online versions. We confined this study to print media because newspapers are the oldest news media in Bangladesh compared to broadcast and online-only media. \textcolor{blue}{To recruit participants in this research, we identified 15 of the most circulated print newspapers in Bangladesh~\cite{bdmedia}. Then we reached out to 30 journalists of these outlets (two from each) via emails and phone calls. We recruited editors and journalists covering political beats. Political beat journalists were recruited because mostly political misinformation spreads to a greater amount \cite{graves2016deciding}. As a result, political journalists need to fact-check beside their regular verification. We also recruited editors because they make editorial decisions.} In total, 12 journalists from nine newspapers- five Bengali-language and four English-language dailies - agreed to give interviews. The Bengali-language newspapers include- \textit{Prothom Alo} (www.prothomalo.com), \textit{Kaler Kantho} (www.kalerkantho.com), \textit{Ittefaq} (www.ittefaq.com.bd), \textit{Jugantor} (www.jugantor.com), and \textit{Manab Zamin} (www.mzamin.com). The English-language dailies include- \textit{The Daily Star} (www.thedailystar.net), \textit{Dhaka Tribune} (www.dhakatribune.com), \textit{The Independent} (www.theindependentbd.com), and \textit{New Age} (www.newagebd.net). Among the 12 interviewees, five were editors, one special correspondent, four chief reporters, and two junior reporters. Among them, two were female and 10 were male.

\subsubsection{Internet Users}
\textcolor{blue}{We conducted an online survey and in-person interviews to understand how Bangladeshi internet users confront, perceive, and combat online misinformation. We designed the survey instrument and interview questionnaire following the findings from existing research on perceived credibility of news \cite{fogg2003prominence, metzger2013credibility} and crowd-source fact-checking \cite{liaw2013maater, nguyen2018believe, hassan2019examining}. The survey included four demography questions, seven multiple choice questions, and one open-ended question. To recruit participant for the survey, we shared the survey link on Facebook. Researchers of this study also requested their friends and families to participate in the survey and share the survey link with others. In addition, the authors shared the survey link to an assorted set of general interest Facebook groups.} In total, we received 500 responses. The participation was voluntary and anonymous. Table \ref{tab-survey-demo} shows demographic (age, education, occupation) information of the participants. About $74\%$ of the participants were from the $21-30$ age range, $78.8\%$ completed Bachelors or a higher degree, and $64\%$ were students. About $63\%$ of the participants were from the capital, Dhaka. 

\begin{table}[h!]
\centering
\caption{Demographic Distribution of the Survey Participants}
\label{tab-survey-demo}
\begin{tabular}{ll||ll||ll}
\toprule
Age      & Freq. & Education  & Freq. & Occupation     & Freq. \\
\midrule
Below 21 & 31    & 8th Grade  & 3     & Student        & 322   \\
21-25    & 243    & 10th Grade & 4     & Housewife      & 2    \\
26-30    & 128    & 12th Grade & 99    & Service Holder & 137    \\
31-35    & 52    & Bachelors  & 238    & Business & 19     \\
36-40    & 27     & Masters    & 142    & Journalist        & 3     \\
Above 40 & 19     & Ph.D.      & 14     & Other    & 16     \\
\bottomrule
\end{tabular}%
\end{table}

To recruit participants for in-person interviews, we followed convenience sampling~\cite{jager2017ii} method to reach out to the respondents. To cover different user groups, we interviewed housemaids, drivers, and elderly citizens in addition to students and service holders. In total, there were 29 participants. They included 13 males and 16 females. They belong to a diverse set of groups- 10 of them were students, 3 car drivers, 4 maids, 5 housewives, 4 service holders, and 3 retired citizens. The distribution of their highest level of education is- 3 with no education, 1 up to 5$^{th}$ grade, 5 in high school, 14 with a Bachelors degree, and 6 with a Masters or above. Table~\ref{tab-user-demo} presents details of the interviewees. \textcolor{blue}{We understand the risks involved with convenience sampling such as bias and insufficient power~\cite{bornstein2013sampling}. However, we argue that our online survey and in-person interview methods complement each other and help reduce the bias involved in each method. Moreover, convenience sampling allows to gather rich qualitative data~\cite{hu2018generalizability}.}

\begin{table}[h]
\caption{Demographic Information of the 29 Users}
\begin{tabular}{l|l|l|l|r}
\toprule
ID  & Age      & Gender & Education  & Occupation     \\ \midrule
D1  & 21 - 25  & Male   & 10th Grade & Driver         \\
D2  & Below 21 & Male   & 10th Grade & Driver         \\
D3  & 31 - 35  & Male   & 12th Grade & Driver         \\ \midrule
H1  & 31 - 35  & Female & Bachelors  & Housewife      \\
H2  & 26 - 30  & Female & Bachelors  & Housewife      \\
H3  & 36 - 40  & Female & Bachelors  & Housewife      \\
H4  & Above 40 & Female & Bachelors  & Housewife      \\
H5  & 36 - 40  & Female & Bachelors  & Housewife      \\ \midrule
M1  & 31 - 35  & Female & 5th grade  & Housemaid      \\
M2  & 26 - 30  & Female & None       & Housemaid      \\
M3  & 36 - 40  & Female & None       & Housemaid      \\
M4  & Above 40 & Female & None       & Housemaid      \\ \midrule
R1  & Above 40 & Male   & Masters    & Retired        \\
R2  & Above 40 & Male   & Masters    & Retired        \\
R3  & Above 40 & Female & Masters    & Retired        \\ \midrule
S1  & 21 - 25  & Female & Bachelors  & Student        \\
S2  & 21 - 25  & Male   & Bachelors  & Student        \\
S3  & Below 21 & Male   & 12th Grade & Student        \\
S4  & 21 - 25  & Female & Bachelors  & Student        \\
S5  & 21 - 25  & Male   & Bachelors  & Student        \\
S6  & 21 - 25  & Female & Bachelors  & Student        \\
S7  & 21 - 25  & Male   & Bachelors  & Student        \\
S8  & 21 - 25  & Female & Bachelors  & Student        \\
S9  & 26 - 30  & Male   & Bachelors  & Student        \\ 
S10 & 21 - 25  & Male   & Bachelors  & Student        \\ \midrule
SH1 & Above 40 & Female & Masters    & Service Holder \\
SH2 & Above 40 & Male   & 12th Grade & Service Holder \\
SH3 & 31 - 35  & Female & Masters    & Service Holder \\
SH4 & Above 40 & Male   & Masters    & Service Holder \\ \bottomrule
\end{tabular}
\label{tab-user-demo}
\end{table}

\textcolor{blue}{\subsection{Interview and Survey Protocol}
We followed a mix of structured and unstructured interview questions. Interviewers followed the predefined questionnaires (see Appendix) to conduct the interviews. Supplementary questions were also asked for more details. After the initial round of the interviews, we reached out to the fact-checkers and journalists again to follow up with some additional questions. Interviews with fact-checkers were conducted via Skype. The interviews were recorded. The average length of the interviews was 80 minutes. Among the interviews with journalists, seven were conducted via Skype and five were in-person, with an average length of 33 minutes. We recorded the interviews. In-person interviews with internet users lasted for an average length of 30 minutes. Consents were taken from the participants following the IRB guidelines. The interviews and the survey were conducted in the native language, Bengali. Two authors of this paper, who also have journalistic experience of an average of five years, conducted the interviews, then transcribed, and translated into English. The two interviewers are native Bengali speakers and proficient in English.}

\textcolor{blue}{\subsection{Data Analysis}
We applied thematic analysis method to analyze the interview data. Following the methodological lessons of Bryman \cite{bryman2016social} on
thematic analysis, two authors coded the interview transcripts of journalists, fact-checkers, and users. The two authors first separately read the transcripts several times to find out themes and its supporting evidence. Specifically, first, the two coders transformed the raw interview data into brief codes. Then, we found connections among these codes following axial coding principles~\cite{scott2017axial}. For example, we identified the internet users' responses that were related to misinformation perception. After that, we analyzed the codes, identified patterns among them, and developed an initial set of themes covering the underlying structure of experiences or processes that were evident in the raw data. We conducted within-narrative and between-narrative comparisons~\cite{heist2012thematic} while developing the themes. Then the two coders together discussed the themes until consensus was reached. To report on interview findings, we kept interviewees anonymous considering their safety.}

\section{Findings}
\label{findings}

\subsection{Mainstream News Media} 
\label{find_j}
The top editors and senior reporters in Bangladesh, interviewed for this study, agreed unanimously that misinformation spreading on the Internet was causing harm to society. But their organizations did not have the capabilities and resources to contain them. Almost none of these news organizations has a fact-checking unit meant to debunk false information floating on the web. Some have social media units that monitor the internet primarily to find news tips. Only two journalists -- one from daily \textit{Prothom Alo} and the other from \textit{The Daily Star} -- were able to recall a few stories their newspapers had published about some viral photographs. The photos were falsely associated with events and got published in some leading newspapers. Only a few interviewees knew about the fact-checking organizations in Bangladesh, and no one ever cited them in their stories. However, these journalists suggested that fact-checking by independent organizations would supplement and improve journalism and hold professional journalists more accountable.

In summary, three major themes emerged from the interviews with the journalists in Bangladesh.
\begin{itemize}
    \item Hands of news organizations are tied as misinformation spreads on the Internet.
    \item News organizations monitor social media to find news tips, not to counter misinformation.
    \item Fact-checking by independent organizations supplements journalism.
\end{itemize}

{\bf Hands Tied As Misinformation Spread.} Although journalists routinely perform verification and possess the skills required for fact-checking, the mainstream news organizations in Bangladesh follow traditional routines and gatekeeping processes. Journalists select stories based on newsworthiness and verify claims made by people with authorities (e.g., government officials and politicians) before adding them to stories. Some interviewees said they would welcome a separate fact-checking unit in their newsrooms, but the objective of such a unit would be to strengthen the verification process as government sources and politicians often make false or misleading claims.

For example, journalist 1 of \textit{Dhaka Tribune} noted: \begin{quote} Political leaders often provide inaccurate information and, in some cases, lies. Recently, we have heard that a conspiracy to kill our prime minister was going on. Political leaders from her party repeatedly said it. But later it was found out to be incorrect.\end{quote}

But an additional fact-checking unit would require additional resources and funding that most organizations cannot afford. Journalist 2 of \textit{New Age} said: 
\begin{quote} It would have been great if we had such a research unit, but it requires human resources and funding. We are facing a crisis in both cases.\end{quote}

However, the interviewees said fact-checking information floating online was not part of a journalist's job. It is beyond their ability given the amount of information being published online and the complexity of the situation. Some suggested that some viral misleading stories drew their attention and they were willing to publish a response to help their readers, but they were unable to do it. Citing the suspension of several Facebook and Twitter accounts that were allegedly run by people with links to the government to spread misinformation, journalist 3 of a Bengali-language daily said intelligence agencies are often behind the spread of misinformation. The journalist noted, \begin{quote} The government started spreading fake news in Bangladesh. We have seen such activities during the recent national election when a bunch of fake or duplicate websites of national and international news outlets were made to spread fake news.\end{quote} Journalist 4 of an English-language daily also noted that intelligence agencies were a major source of misleading information. While other journalists refrained from accusing any government agency of spreading false information, they unanimously agreed that journalists in Bangladesh lack freedom to tell stories in a neutral manner.

{\bf Social Media Usage Not Meant to Counter Misinformation.}
It is evident in the interviews that journalists in Bangladesh keep an eye on social media. While a few news organizations have teams designated for social media only, others rely mostly on reporters. However, the main objective of monitoring social media is to find news tips. Journalist 5 of \textit{The Daily Star} said, \begin{quote} I think social media help us by providing clues to news items, especially when something happens in remote areas of the country.\end{quote} These clues often materialize into news stories and get published if the reporters are able to verify the authenticity of related information. Most of the journalists interviewed echoed this sentiment. Journalist 2 of \textit{New Age} said, \begin{quote} In some cases we find something newsworthy on verified pages of socially, culturally, politically or nationally important persons...we consider them news.\end{quote}

For most news organizations, uses of social media are not generally meant to fact-check misinformation or write pieces debunking misleading news. However, some journalists said they occasionally write stories about misleading photos if they get the necessary evidence to prove they are misleading. The examples they provided indicate that they would write such stories if those stories get published in some mainstream news media. For example, journalist 5 of \textit{The Daily Star} recalled running a fact-checking story. \begin{quote} During Rohingya crisis, a mainstream newspaper in Bangladesh published some pictures which were fake. Some of the pictures were from Nigeria. We published a counter piece describing that content the newspaper published was fake. We did it because we thought such fake items would interrupt social and religious harmony in Bangladesh.\end{quote}

One key concern that emerged from interviews with journalists in Bangladesh is that misinformation often gets published in known newspapers as they rely on social media for news. Most editors, however, claim that they do not publish unverified claims or information from social media. Journalist 6 of \textit{Jugantor} explained the process of dealing with misleading information online. \begin{quote} When something viral appeared to be a rumor, our news management team works on the issue and verifies credibility of the information. Me and a few other news editors, joint news editors, associate editors, chief reporter and shift in-charges work in the team.\end{quote}

Yet, some rumors get published and journalists agree that misinformation published in newspapers can be more damaging than misinformation spreading on the Internet. Journalist 7 of \textit{Prothom Alo} claimed: \begin{quote} We have seen some online news outlets publish news in a hurry without proper verification. People share these pieces on social media. Mainstream newspapers sometimes publish the same items without further verification.\end{quote} 

However, journalist 7, whose organization is one of the few that does fact-checking, said the goal of his social media team is to make sure that misleading information does not get published in his newspaper, but not to counter misinformation found on social media.

{\bf Independent Fact-checking A Growing Avenue of Journalism.} Most of the interviewees agree that fact-checking by independent organizations would improve the quality of journalism given that they follow the accuracy standards of journalism. For instance, journalist 8 of \textit{Kaler Kantho} said, \begin{quote} I think fact-checkers will make journalists and news media more accountable... Their critical eyes on our work would make us more responsible.\end{quote} Journalist 9 of \textit{Manab Zamin} strongly emphasized on the need for independent and unbiased fact-checking organizations. Journalist 2 agreed: \begin{quote} Fact findings supplement journalism... I would have welcomed them. But it is hard to assess their reliability and integrity. If they maintain these, they are welcome. They might make journalists' work easy.\end{quote}

Some argue that fact-checkers perform only a part of journalism. In other words, fact-checkers are carving out a niche within the field of journalism. Journalist 10 of \textit{Dhaka Tribune} suggested that fact-checkers take on some tasks that journalists working for mainstream news organizations find hard to do. For instance, \begin{quote} reporters are not well equipped with technology and technical issues, in which fact-checkers are doing well.\end{quote} A few other journalists such as journalist 7 echoed the sentiment. Journalist 10 also suggested that fact-checkers may deal with rumors and misinformation floating online while news organizations can stick to their typical news routines.

However, journalists in Bangladesh are divided on whether they need a separate fact-checking unit in their own organizations. Some said a specialized fact-checking unit would help them while others think reporters are already doing fact-checking. 

\subsection{Fact-checking Organizations}
\label{subsec-fco}

Fact-checkers in Bangladesh, interviewed for this study, noted that they started verification in response to the increasing amount of misinformation and rumors on social media. Most of these initiatives such as Jaachai and BDFactCheck were started voluntarily by some university students and some young journalists. By contrast, Fact Watch began as an initiative of the Media Studies and Journalism department of the University of Liberal Arts Bangladesh, receiving a grant from the American Center in Bangladesh. Overall, the motivation for setting up fact-checking initiatives in Bangladesh was to debunk misinformation that went viral on the web and to offer facts to the general public who is most likely susceptible to misinformation. Despite various constraints such as unavailability of data archives, political pressure, and scarce human and technological resources, these initiatives have been verifying claims relating to politics, health, religion, and consumer services.

The following themes emerged from interviews with seven fact-checkers from three organizations.

\begin{itemize}
    \item Working voluntarily interferes with paying full attention to fact-checking.
    \item Political pressure and scarce resource lead to biased selection of topics for verification.
    \item Fact-checkers mostly use information available in open sources to verify claims.
    \item Fact-checkers are lagging behind due to the lack of technological support. 
\end{itemize}

\textbf{Working Voluntarily Interferes With Fact-checkers' Work.} The fact-checkers in Bangladesh have started the work of fact-checking from their motivation to offer facts to the general public, without having any prior training and institutional infrastructure such as an office space and payment for work. Most of them perform this work in addition to their full-time job and studentship. This results in interfering with paying full attention to their work. For instance, fact-checker 1 of BDFactCheck said, ``I wish I could take it as my main profession. But I have to do my main job to earn my living. I am doing fact-checking from my passion.'' The team of Jaachai paused their work in December 2018 because the members could not put in enough time to do verification in addition to their regular jobs.

These initiatives have been functioning voluntarily because they cannot apply for any grant. They would need accreditation from the government to apply for a grant; however, the government of Bangladesh does not have any procedure to offer an accreditation to fact-checking organizations. Fact-checker 1 noted: \begin{quote} The primary requirement to apply for any grant is to have registration, which we do not have. We also cannot apply to be registered as a fact-checking organization with IFCN for that.\end{quote} The only exception was Fact Watch that started its work as a platform of the University of Liberal Arts Bangladesh, receiving a ten-month grant from the American Center. During that period, its fact-checkers got paid. However, with the expiration of the grant, Fact Watch was in transition to be introduced as a co-curricular activity of the university.

It is evident in the interviews that fact-checkers in Bangladesh started the work of verification without having any prior training. The work of western fact-checkers has been an inspiration for them. For example, given the growing amount of misleading claims on social media fact-checkers with Jaachai started verification of those claims and then published those on its Facebook page named ``Share content with utmost care''. Fact-checker 2 of Jaachai said: \begin{quote}We had little idea about the term ``fact-checking'' when we first started debunking claims. Later, we got to know about this practice looking through the work of \textit{Snopes} which inspired us to turn our Facebook page to ``Jaachai'' with its own website.\end{quote}

\textbf{Political Pressure And Scarce Resource Lead To Biased Selection Of Topics For Verification.} One key concern that emerged from the interviews was that political pressure interferes with fact-checkers' selection of topics for verification. Fact Watch has decided to verify claims relating to public health and environment only, not political claims, fearing political backlash, ramifications, etc. Fact-checker 3 of Fact Watch explained:

\begin{quote} Some misinformation goes against the government; some, on the other hand, support and praise the government. Although the government wants to combat misinformation, but not always when the items support its activities. This practice creates a problem for independent fact-checkers. We cannot choose all the claims for verification with absolute freedom. Such situations lead us to be self-censored.\end{quote}

Fact-checker 4 of BDFactCheck echoed similar concerns. The fact-checker said, ``When we choose political claims for verification, we have to keep in mind whether or not it would be very critical of the government.''

Fact-checker 1 provided an example:

\begin{quote} We were worried when we chose to verify the claim whether the prime minister of Bangladesh was enlisted as the second most honest leader across the world. The claim went viral on social media. We took the risk and verified the claim and found that it was false. Later, an official of a government agency warned us not to do such report; otherwise our website will be blocked. \end{quote}

Scarce resources and lack of expertise on certain topics interfere with topic selection. For example, fact-checkers at BDFactCheck cannot always verify claims relating to health. Fact-checker 4 said:

\begin{quote} A reader requested us to verify whether piercing any part of body with a needle can save a person suffering from a heart attack. But we could not do it because we do not have enough knowledge of medical science. We also could not reach out to any doctors. Doctors usually do not respond to our queries.\end{quote}

However, Jaachai team got assistance from its readers in its Facebook group to verify health-related claims. ``There are some doctors in our group. They help us with medical explanations,'' said fact-checker 2. 

Although due to the resource constraint the fact-checkers of three organizations cannot put enough time to monitor social media for locating viral claims, readers of these organizations' Facebook groups help them locate viral claims on the web and provide evidence relating to those claims.  

\textbf{Fact-checkers Mostly Use Information Found In Open Sources For Verifying Claims.} It is evident in the interviews that fact-checkers in Bangladesh mostly used information available online. Several reasons such as fear of disclosing identity and not having access to the primary sources are responsible for this practice. Since the beginning Jaachai's editorial policy was not to disclose identity of its team members. This policy interferes with topic and source selection. Fact-checker 2 of Jaachai said, ``As we maintain anonymity, we use information available online.'' 

The interviewees said they google to find relevant information. Fact-checker 5 of BDFact-Check said:

\begin{quote} I checked a photo published in a Bengali-language newspaper that said a flock of birds died due to radio waves emitting from mobile phone towers in the Netherlands. First, I checked the photo in Google image and found that the context was different: the birds died of chemical contamination. Then I checked where else the piece was published and found that it was published in some fake news sites. I did not find the piece in any reliable media in the Netherlands. So, I decided the photo was fake.\end{quote}

The team members of Fact Watch are university students. They face problems with access to both primary and secondary sources. To resolve this issue, Fact Watch recently has signed an MoU (Memorandum of Understanding) with an English-language newspaper in Bangladesh, in which its fact-checkers would seek help from the journalists of the newspaper to get access to public offices. However, fact-checkers with BDFactCheck contact primary sources when needed. For example, fact-checker 4 said, 

\begin{quote} An online news portal in Bangladesh published a report claiming the prime minister of Bangladesh as the Asian Mandela. The portal quoted an Australian professor. We emailed the professor to find out whether he said this. Two days later, he replied that he did not say anything like that.\end{quote}

\textbf{Fact-checkers Are Lagging Behind Due To The Lack Of Technological Support.} It is found in the interviews that lack of technological support is lagging behind the fact-checkers in Bangladesh. For example, fact-checker 2 said, 

\begin{quote} An image of floating dead bodies in a lake during safe road movement in Bangladesh went viral on Facebook. We applied keyword search on Facebook, a time-consuming approach to find out the original source. It took several hours to find its original source using the keyword ``lake''. In the meantime, the fake post might have reached millions of users.\end{quote}

The fact-checker also noted, ``When we do searches on Google, fake sites also pop up. It is really difficult which sites we would consider as authentic, a big challenge for us.''

A lack of data archives in Bangladesh has been a persistent problem for fact-checkers to do verification. Fact-checker 4 noted,

\begin{quote} When we collect data for checking a claim, we can't find it because Bangladesh does not have any data storage facility. We depend on crowd source to gather information; for example, we have a Facebook group where we put some topic to gather ideas and facts from them.\end{quote}

Moreover, most of the fact-checkers in Bangladesh are not aware of the software they can use to do verification. Fact-checker 6 of Fact Watch said, ``Although my university is willing to pay money to buy software, but I do not have knowledge about the tools we could use.'' 

However, fact-checker 4 said,

\begin {quote} Fact-checkers in the West are using automated tools for verification. But there is no such tool in Bengali language. Western fact-checking organizations are also using machine learning and artificial intelligence to identify whether a statement is fake or real. We do not have such technological support.\end{quote}

\subsection{Fact-checking by General Users}
To understand how internet users in Bangladesh are combating misinformation, we interviewed a group of $29$ diverse people in-person and conducted an online survey with $500$ individuals. When we put the findings from these two studies side by side, some patterns become apparent. Below, we describe i) the extent of misinformation experienced by the internet users, ii) different combating strategies taken by them, and iii) what measures may work in limiting misinformation according to their perspectives.

\begin{figure}[h]
\centering
\begin{minipage}{.5\textwidth}
  \centering
  \includegraphics[width=\linewidth]{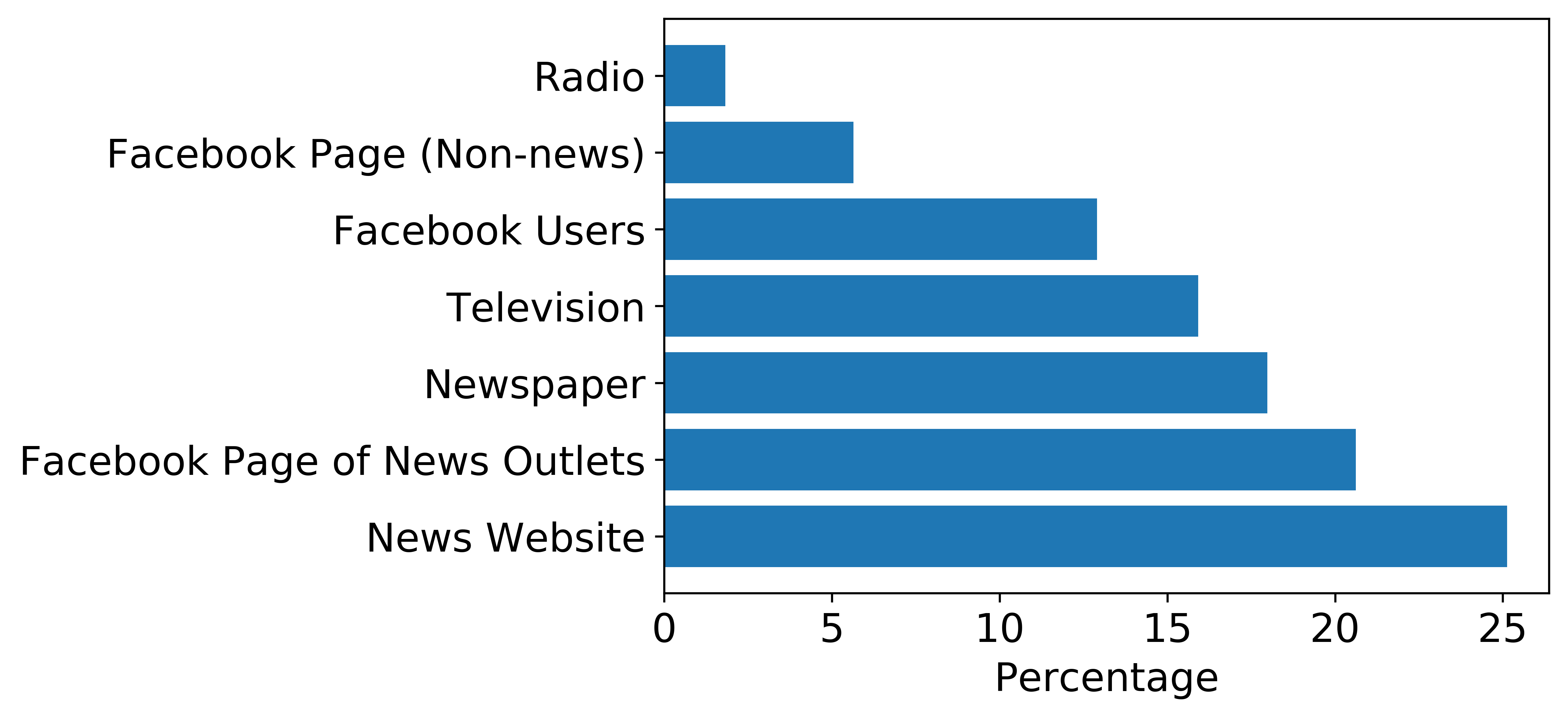}
    \caption{Source of Getting News}
    \label{fig:news_source}
\end{minipage}%
\begin{minipage}{.5\textwidth}
  \centering
  \includegraphics[width=\linewidth]{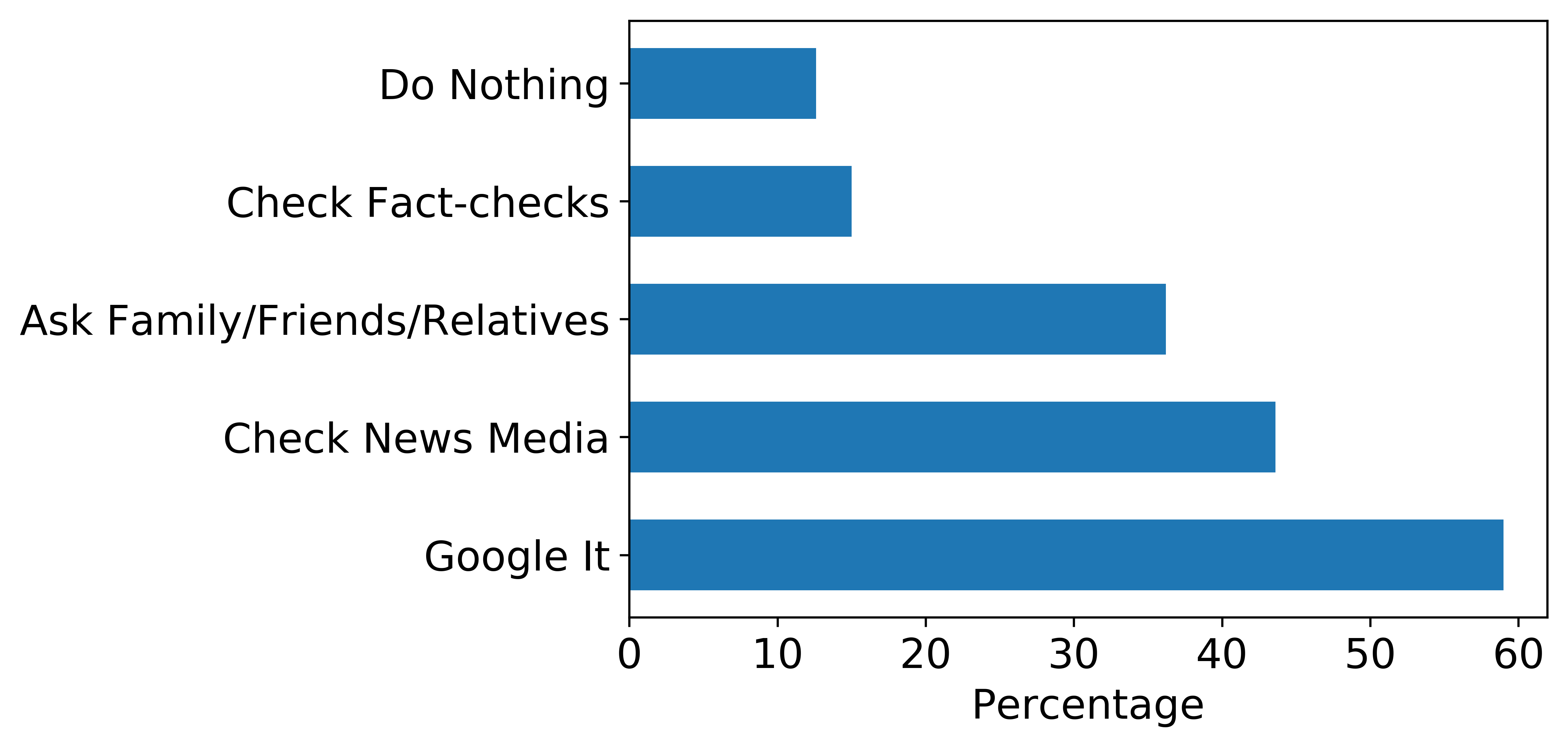}
    \caption{Steps Taken if Credibility is in Question}
    \label{fig:measures}
\end{minipage}

\end{figure}

\textbf{Fake News is Everywhere. Or \textit{is it?}}
First, we aimed to understand the ways through which internet users of Bangladesh consume news. Figure~\ref{fig:news_source} shows different sources of getting news. In the survey, we asked the users to select the sources from which they consume news. If they consume news from multiple sources, they can select multiple options. They can also add sources which are not provided as an option. Figure~\ref{fig:news_source} shows that about 400 survey participants (25.13\%) selected ``News Website'' as their news source. People also follow popular Facebook users/celebrities to get news. About 100 participants (20.61\%) stated that they consume news from Facebook pages which don't belong to a news portal. The common sources which were not in our provided options but later added by the users are- YouTube (2) and Twitter (3). We asked two boolean questions in the survey- \textit{i)} whether or not the participants saw any misinformation in their social news feed and \textit{ii)} they ever found a piece of news to be false which they initially thought to be true. $96.2\%$ and $90.6\%$ of the participants responded YES to the first question and second question, respectively. The first number indicates the prevalence of the misinformation problem. The second number says something more. It indicates that a large number of people don't realize if a news story is false in their first encounter but later found the news to be false. When we control for the age factor with these questions, we see that all participants below $21$ reported to have seen misinformation in their social media feed. 

\textbf{Fact-checking Strategies.}
One survey question asks what users do if they have any doubt about the credibility of a news story. Figure \ref{fig:measures} shows the distribution of measures the users generally take if they are in doubt about the credibility of a news story. If a user generally considers multiple measures, she/he could select multiple options. That is why, the sum of all percentages is more than 100. Overall, the most common step the users take is they go to Google.com to check other resources. This is particularly common among users with age below 21. About $40\%$ of them search Google.com. The third most common step, about $10\%$, taken by this age group is checking other news media. On the other hand, among the respondents with age more than $40$ years, $33\%$ of them reported to check other news media when they are in doubt; this is the most common strategy among this age group. The fact-checking organizations have had some limited success in gaining popularity among the users. About $15\%$ of the overall survey participants mentioned that they seek information from one of the fact-checking organizations when they have doubt about a news story. Another common strategy is asking friends and family members for further verification. This strategy is more common among young users with age below $21$ ($32\%$, second most common strategy) compared to respondents with age above $40$ ($7\%$, least common strategy).

\textbf{Responsibility of Fact-checking.} Users have a difference in opinion about who should be responsible for keeping the online (environment, landscape, etc) safe from misinformation. 62.0\% of all the 500 survey participants think that it is the responsibility of the \textit{Social Media Platforms} (e.g, Facebook, Twitter) to combat online misinformation. 57\% think \textit{Journalists} are responsible, 46\% think of \textit{Fact-checkers}. 43\% of the users think that people should be informed about online misinformation and educated about how to identify them (\textit{Train People}). Table \ref{tab-responsible} shows users perception, from different age groups, about who should be responsible for combating online misinformation. We find that 26.1\% of the users (the most) with age below 21 think \textit{Social Media Platforms} are responsible and only 15.9\% (the least) think people should be trained. We observe a completely opposite view from the users with age above 40. Most of them (22.4\%) think people should be trained and 17.2\% (the least) put the responsibility on \textit{Social Media Platforms}.\\

\begin{table}[]
\caption{Users' Perception of Misinformation Combating Responsibility. Numbers Indicate Percentages.}
\begin{tabular}{l|l|l|l|l|r}
\toprule
Age & Journalists & Fact-checkers & Government & Social Media & Train People\\ \midrule
Below 21                                        & 17.0                                                                  & 19.3                                                                    & 21.6                                                              & 26.1                                   & 15.9       \\
21-25                                           & 21.3                                                                  & 16.5                                                                    & 21.1                                                              & 24.4                                   & 16.5                                                                                                                                            \\
26-40                                           & 23.69                                                                  & 17.09                                                                    & 19.81                                                              & 23.11                                   & 16.32                                                                                                                                            \\
Above 40                                    & 20.7                                                                  & 19.0                                                                    & 20.7                                                              & 17.2                                   & 22.4                                                                                                                                            \\
                                                                                                                                
\bottomrule
\end{tabular}
\label{tab-responsible}
\end{table}

The in-depth interviews with 29 users that included people from various socio-economic status provide further understanding about their perception of misinformation and how they deal with such information online. These users are aware that misinformation is being spread through the Internet, and each user has her/his own set of mechanisms to counter what they perceive to be misinformation. Of them, four mechanisms appeared to be more prominent than others.

\textbf{Rely on Higher Authority.} The majority of the users said they would trust information if it had come from or been shared by certain individuals. In other words, these individuals play the role of opinion leaders. For example, the interviewee D3, who is a chauffeur by profession said:

\begin{quote} {Yes, fake news may exist, but I don't really know which ones are fake. So if I see my boss (owner of the car) sharing any news, I think it's credible and I follow it.}\end{quote}

Similarly, we observe that users depend on the more experienced online users for the news verification process. For instance, an elderly retired interviewee R2 noted:

\begin{quote} {My granddaughter is efficient about the Facebook stuff and all these. She is an educated person. So whenever I doubt anything, I just ask her whether I should share/believe that or not.}\end{quote}

A housewife H1 noted that she had shared posts coming from her husband and her sister. As she pointed out: 

\begin{quote} {...also, I trust my sister who is a government official and much educated, so I re-share her posts too. They cannot share anything bad.}\end{quote}

Several other interviewees said they usually follow social media pages of the news organizations which they trust.

\textbf{Share Harmless Information.} Some users believe that a piece of news with a humor tone, even if false or misleading, is mostly harmless and think that it is okay to share them with people for fun. Sometimes, when these users see posts containing funny information, they share them without hesitation. For instance, one user D3 mentioned:

\begin{quote} {If something is funny, then I share it with my friends on messenger, sometimes post on my Facebook page. I don't know if they're fake, even if it be, I don't recheck.}\end {quote}

We find that there is a lack of knowledge among some users about the consequences of misinformation. Some users say that they do not believe that misinformation even relating to serious issues can cause much harm to society. One user, D1 noted:

\begin{quote} {When a violence occurs, it's because of some internal conflicts between people. Only a fake piece of information alone cannot create it.}\end {quote}

Users, especially the elderly ones and the ones in a lower income group, appeared to be naive about the intention of the people who spread misinformation when it comes to topics such as health and religion. They find it hard to believe that people can share news related to health and religion with a bad intention. So, when they find such posts, they share with others without any hesitation. For instance, one user, M3 said:

\begin{quote} {How can people share misleading information about health issues? This is so sensitive and people who share these types of news, cannot be so bad to do that.}\end{quote}

Another user H4 noted, 

\begin{quote} {Well, I don't care if it's fake or real when I'm clear to myself that I have no bad intention. Like, sharing health tips is not bad, since I'm doing for others' betterment and no one would get harmed for this.}\end{quote}

Some of these users also believe that religious posts cannot be fake either. The following response from D2, for example, suggests that it is imperative to believe anything religious without doubt:

\begin{quote} {Religious news is hard to believe as fake. I regularly post these on my Facebook page and our group. If you put any doubt on these religious issues, that means you're doubting our religion. right?} \end{quote}

\textbf{Confirmation Bias.} Many of the interviewed users indicated that one key mechanism which they use to verify misinformation is their own perception and understanding of various issues and events. People with strong religious and political views are more likely to use this mechanism. Some users suggested that they would never share anything that goes against their religious and political views. On the other hand, they would follow and share anything if it supports her or his views. For instance, one user H2 pointed out: 

\begin{quote} Personally I don't share anything that goes against my religious/political views. If it aligns with my views, I really don't care to check the credibility. Anything in favor of my political party, I'll share it for the better publicity of my party.\end{quote}

Similarly, several other users said they would follow individuals or pages if posts on those pages reflect their views.

This phenomenon often goes beyond politics and religion. It is also embroiled in local culture or lack of knowledge. There are cases in which users would consider proven facts as misleading. It is evident in one comment of D2: 

\begin {quote} I saw a post many years ago on Facebook, people from earth has landed on the moon. Well, that's never ever possible. We do not have the power to do it, and even if we do, how can you take a picture on the moon and then send it to earth? \end{quote}

Another response that is associated with local culture is as follows: 

\begin {quote} DBC News [a television station] has shared a piece of news about the marriage of two girls. It cannot be real. A girl cannot get married with another girl! This is absurd.\end{quote}

\textbf{Verification.} There is a small number of users, most of them are students at local universities, who say they verify information that they find on social media. One student S3 explained how he verifies information: 

\begin {quote} When I see any political post and have doubt on that, I check who shared this. If it's from a fake account, then I don't trust it. If it's from a locked account, then I check the mutual friends. If I have no mutual friends, then the doubt gets stronger. But if the mutual friends are whom I trust, then the doubt weakens.\end {quote}

According to this student, accounts spreading false information usually use photos that are downloaded from the Internet. Such accounts also have little information about themselves in their profile page. 

Another student S8 suggested that it is important to develop analytical ability and think logically to be able to stay protected from misinformation.

Although these mechanisms appeared to be dominant in the interviews with users, a few other important points emerged from the interviews. For example, some users say that they avoid sharing posts on controversial topics. Most of them emphasized that they would welcome sites that would verify facts for them. Finally, students seemed to be better aware and prepared to deal with misinformation than users in other groups.


\section{Discussion}
\label{discussion}
In this study, we have examined how multiple entities - journalists, fact-checkers, and internet users - in Bangladesh have experienced and confronted online misinformation with limited resource, freedom of press, and access to information. We have reported the communication gap between journalists and fact-checkers that resulted in a lack of defined responsibility of who should fact-check online misinformation. We also have identified several verification strategies of regular users. Our findings provide several design and policy implications, \textcolor{blue}{and suggest the need for development of computational tools and increasing cooperative work to combat misinformation in the developing world. Below, we discuss the findings in light of the context of Bangladesh, suggest several broader implications in terms of design and policy in the context of similar socioeconomic regions and beyond, and recommend several research directions for the CSCW/CHI community.}

\subsection{Combat Misinformation in Bangladesh}
\subsubsection{Free the Press}
Our interviews with journalists and fact-checkers suggest that freedom of the press needs to be improved to facilitate the fact-checking practice. The Bangladesh government has enacted laws to curtail misinformation (see section \ref{ict}); however, our findings suggest that these laws are not enforced equally. Any verification that goes against the government might put journalists and fact-checkers in danger. As a result, they cannot verify all misinformation freely. \textcolor{blue}{Political pressure is leading to a biased selection of fact-checks (see section~\ref{find_j} and \ref{subsec-fco}).} Moreover, intelligence agencies, as stated by our interviewees, are spreading misinformation. The government also has a custom-made judicial weapon for silencing troublesome journalists – the 2018 digital security law, under which ``negative propaganda'' is punishable by up to 14 years in prison. As a result, self-censorship has reached unprecedented levels because editors are reluctant to risk imprisonment or their media outlet’s closure \cite{reporters2020borders}. The government of Bangladesh needs to enforce laws equally, take measures to free the press so that mass people get accurate information. \textcolor{blue}{CSCW researchers have long been studying the use, appropriation, and manipulations of media platforms by powerful entities like the government~\cite{trottier2014social, bradshaw2017troops}. Several recent studies in this area have also shown how political leaders are abusing social media platform by spreading misinformation and hate speeches. Hence, there needs to be a more strict guidelines and democratic checking on social media to regulate the misinformation related campaigns~\cite{gillespie2010politics, pal2015banalities, pal2016twitter}. }

\subsubsection{Collaboration Between Journalists and Independent Fact-checkers}
The key objective of this study was to have a deep understanding of people's perception regarding whose responsibility it is to combat misinformation in Bangladesh. Findings from in-depth interviews with editors and senior journalists suggest that the journalists at mainstream news media are not equipped to take on this responsibility. This results from the lack of resources and fear of repercussions from the government. A majority of the interviewees suggested that the task of combating misinformation should be performed by independent fact-checking organizations. On the other hand, fact-checkers suggested that they had to take on this moral responsibility because no one was responding to the growing amount of misinformation online. As they do not have any institutional structure, journalists from the mainstream media do not know about them. Fact-checkers also cannot collaborate with the journalists due mainly to two reasons. First, most of the journalists do not know about their work. As a result, fact-checkers' work do not circulate through mainstream news media. Second, the journalists who know about fact-checkers' work do not cite or publish in their newspapers because they cannot rely on their verification. While in the U.S. independent fact-checkers, such as \textit{FactCheck.org} and \textit{PolitiFact} have collaborated with national and local news organizations to disseminate their work and assist journalists with fact-checking \cite{amazeen2013making}, such practice is absent in Bangladesh. Fact-checkers in Bangladesh could consider the collaboration with mainstream news organizations to share their work with a broader audience. Our findings from the users indicate that they want the mainstream news organizations to verify online misinformation and present to them. As journalists with mainstream newspapers in Bangladesh cannot do that due in part to resource constraint, they could also consider collaboration with independent fact-checkers, which eventually would address the problem of misinformation in Bangladesh. \textcolor{blue}{CSCW researchers have long been interested in scientific collaboration, and a version of that might be applied to develop a global network of fact-checking. We argue that such transnational collaboration is important for limiting the spread of misinformation over social media, because of the ubiquitous nature of the platform ~\cite{velden2013explaining,velden2014sharing}.} 

\subsubsection{Collaborating with Faith-based Institutions}
Our findings show that users in Bangladesh tend to believe in online religious content. While social media platforms such as Facebook have been developing computational (based on natural language processing and machine learning) techniques to detect religious fake news and hate posts, we suggest a Human Computer Interaction (HCI) approach involving the faith-based institutions. \textcolor{blue}{Bangladesh is a Muslim-majority country with at least 90\% Muslim population. It has 250,000 mosques, about 4 mosques per square mile \cite{bdnews2011mosques}. The local Imams (religious leaders) are usually very respected in the community and they often play the role of opinion leaders. These Imams are also connected with each other through an online platform which is made available by the government \cite{bd2020imam}. Facebook and other social media platforms can collaborate with these institutions to detect and correct locally generated religious misinformation. Bangladeshi fact-checkers also could interview religious leaders in their verification process to make people aware that what they see online are not always necessarily true. The relationship between faith and computing is getting growing interest in CSCW and related disciplines (see ~\cite{rifat2017money, sultana2019witchcraft, mim2020others}, for example). While most of the work in this area is highlighting the differences between the core assumptions between mainstream computing and traditional religions, and associated challenges, we argue that a right combination of these two is both possible and necessary. A growing body of work in the Global South has shown how motivation for social good can be harnessed through the help of religious institutions~\cite{batool2019money, rifat2017money}. We believe that much more research is needed in CSCW and related disciplines to find the right method to engage religious values to address societal problems like misinformation~\cite{sultana2019witchcraft, sultana2020parareligious, mustafa2020islamichci, ibtasam2019my}.} 

\subsubsection{Access to Information}
The interview results showed that fact-checkers do not have adequate access to information. This is mainly due to the cultural reasons and lack of digital archives. Official Secrecy Acts of British era are still in effect in Bangladesh \cite{laws2020bangladesh}, which resulted in a culture of secrecy. Although Bangladesh has enacted the Right to Information Act in 2008 which enables people to seek public information, in many cases, the Act does not help journalists due to the long period of time it takes to get information \cite{rti2019bangladesh}. As the Bangladesh government envisions to offer public services digitally and develop the country incorporating digital technologies, the government needs to take an open data policy in which they could offer information through digital archives. The government also could take initiatives of making people media literate by introducing media literacy curriculum at schools and colleges. In addition, the government could use mass media and folk media to spread messages relating to media literacy.

\subsubsection{Technology Design}
We observe that most of the automated fact-checking efforts depend on advanced technologies such as knowledge base, computational linguistic resources, and computation and data experts in the newsrooms. We suggest that development of linguistic resources for Bengali language such as parts-of-speech tagger, entity recognition system would facilitate automated fact-checking for news in Bengali language. \textcolor{blue}{Making computing more accessible for all by incorporating non-English languages has been a long demand of CSCW community, and a plethora of cross-cultural communication research has supported this cause by providing various computational and human-mediated tools and techniques~\cite{10.1145/2631488.2634061, 10.1145/1518701.1518806, 10.1145/1031607.1031712}. Additionally, the Global South research within CSCW is also focusing on such cultural adaptation of western technologies for a long time. This paper joins that thread of CSCW research, and calls for developing computing technologies that can understand local languages with their cultural nuances and thus help fact-checking in non-English contexts~\cite{kayan2006cultural, lim2017making}.}

\subsection{Broader Implications}
\textcolor{blue}{Our findings have several broader implications in terms of design and policy recommendations for geographical regions that are similar to Bangladesh with respect to socioeconomic structure and ideological belief systems. For instance, we discussed religious misinformation that is prevalent in countries like India and Myanmar (see section~\ref{misinfoaglobalchallenge}). One of the common dimensions among these countries and Bangladesh is- single major religion. At least 80\% of their citizens follow a single religion- Bangladesh: Islam 90\%, India: Hinduism 80\%, Myanmar: Buddhism 88\%. In all these countries, religious minorities are often being the target of religious misinformation. We argue that effective and persuasive misinformation combating strategies can be developed in these countries and similar other regions with single major religion (e.g., middle east) by designing close collaborative communication between fact-checkers and faith-based institutions. These countries also have similar socio-economic structure (developing economies) \cite{un2014economic}, literacy rate (rank within 126 -- 129) \cite{development2020goals}, and press freedom index (rank within 139 -- 151) \cite{rsf2020ranking}. An increased effort through policy changes and actions to advance press freedom and access to information by the governments of these countries and other regions with similar attributes (e.g., Africa) can facilitate nonpartisan fact-checking and thereby reduce misinformation.}








\subsection{Limitations}
\textcolor{blue}{This research has some limitations. First, the survey samples were not nationally representative as it was conducted through Facebook. The samples included mostly educated, young Facebook users. As a result, those who were not on Facebook were excluded in the survey. To fill this gap, we conducted in-depth interviews with 29 ordinary citizens to include participants from different age groups and education levels. These respondents were included using convenience sampling. Thus, these samples were not nationally representative.}

Overall, this exploratory study reveals new ways to examine this problem for further insights. Clearly, this can be studied from the media management and economic perspectives as fact-checking organizations all over the world have been struggling to find sustainable business models. From the normative perspective, the spread of misinformation and efforts to contain it raise countless new questions about the roles of news media in developing democracies, which need to be answered. \textcolor{blue}{We believe that our study makes important contribution to advance CSCW scholarship in infrastructural politics,  global development, and post-secular and postcolonial computing to address these issues. }

\section*{Acknowledgements} 
We thank the anonymous reviewers for their suggestions and criticisms. We also wholeheartedly thank Meaghan Knight, Cody Buntain, and Hanuma Teja Maddali for their valuable feedback on the draft of this article. The generosity of fact-checkers in Bangladesh helped us explore the new professional world of fact-checking. We owe special thanks to the fact-checkers of BDFactCheck, Fact Watch, and Jaachai. We also thank the journalists for their valuable insights into the ecology of misinformation in Bangladesh. 

This research was made possible by the generous grants from Natural Sciences and Engineering Research Council (\#RGPIN-2018-0), Social Sciences and Humanities Research Council (\#892191082), Canada Foundation for Innovation (\#37608), Ontario Ministry of Research and Innovation (\#37608), and International Fulbright Centennial Fellowship of Syed Ishtiaque Ahmed.



\bibliographystyle{ACM-Reference-Format}
\bibliography{main}
\received{January 2020} 
\received[revised]{June 2020}
\received[accepted]{July 2020} 

\newpage
\appendix

\section{Appendix}
\subsection{Interview Questionnaire for Users}
\begin{enumerate}
    \item Which media do you use to get news?
    \item Have you ever seen any fake news in your news feed?
    \item When in doubt about the authenticity of any of the information, what do you do?
    \item You thought a piece of news to be authentic but later found that it was false. Has this ever happened to you?
    \item From which source did you get that news?
    \item Did you see that in a group? Which one? Can you please elaborate?
    \item Has it ever happened to you that at first, you shared the news/post and then it came out to be false? What did you do then? Did you remove that?
    \item How did other people react to that incident?
    \item What can be done to prevent the spread of false news, especially in Bangladesh?
    \item Are you aware of any fact-checking organization in Bangladesh? Who are they?
    \item According to your opinion, what can be done to verify the authenticity of the news accurately?
    \item \textbf{Demographic Information: } We will collect some demographic information about you, which will not include any identifiable information. This information will give us a better understanding of general demography of the respondents. Can you please answer these questions: What is your age range? How do you identify yourself gender-wise? What is your highest level of education? What is your occupation? 
\end{enumerate}

\subsection{Interview Questionnaire for Fact-checkers}

\subsubsection{Lead-in}
\begin{enumerate}
\item When did you run the organization?
\item Does your organization have own website and/or social media pages?
\end{enumerate}

\subsubsection{Team}

\begin{enumerate}
\item How many members do you have in your team?
\item How did you find the team members?
\item What did motivate you and your team members in fact-checking? Please share any stories behind your motivation.
\item Do you have an office or a place to meet?
\item If no, how do you communicate with team members?
\item If yes, how do you pay rent?
\item Do your team members get paid or do they work voluntarily?
\item If team members work voluntarily, what are their actual professions?
\item Do you and team members have any prior training/experience on fact-checking? If yes, where did you get it?
\item How do you distribute work responsibilities among team members? Please share examples.
\item Can team members give full attention to their job, if working voluntarily? 
\item How many years of journalistic experience do you have? How about other members?
\end{enumerate}

\subsubsection{Organization}
\begin{enumerate}
\item Does your organization have government accreditation?
\item If no, have you ever applied for an approval?
\item Does the Government of Bangladesh have any mechanism to give an approval to a fact-checking initiative?
\item If you do not have government accreditation, what types of problems do you face in doing your job? Please give examples.
\item Follow up with specific job type- gather info from govt. or non-govt. entity? Gather info from other media entities? Contact donors for fund?
\item Does your organization have any funding? If yes, who offered the funding? Does the donor give any instructions what to cover and what not? Please give examples.
\item If you do not have funding, have you ever applied for a funding?
\item Do you think fact-checking initiatives should run for profit or non-profit? Why?
\end{enumerate}

\subsubsection{Topic selection}
\begin{enumerate}
\item Which topics do you verify? 
\item How do you select a topic for verification? Please describe the process with examples.
\item Do you feel that you are free to choose any topic for verification? Why do you choose certain topics? For example, you choose a health-related topic, why? Please give examples.  
\item What are the topics you cannot choose and why? Please give examples.  
\item Do you have prior and adequate expertise on the topics you verify? For example, when you verify a science or health related topic, how do you verify it?
\end{enumerate}

\subsubsection{Sources}
\begin{enumerate}
\item What are the sources you use to verify a claim? Please give examples of using both primary and secondary sources.
\item How do you find the sources? 
\item How do you find relevant sources (both primary and secondary)? 
\item Do you have any database of sources?
\item What are the secondary sources you use? How do you get access to it? Please give examples.
\item Do you mainly use the documents available on the Internet or you visit in person the offices where information can potentially be found?
\item Suppose you need a document from a government office. How do you get access to it?
\item How do you introduce yourselves while contacting sources? Do you experience any constraints in reaching out the sources since the idea of fact-checking is relatively new in Bangladesh? Do they respond when asked for cross-checking information?
\item How do you get access to government officials? Do you get your required information from them? Please share any examples.
\item Do you take help from journalists to reach out to a source? Please share examples. \item A source might spoon-feed you. How do you assess source credibility?
\end{enumerate}

\subsubsection{Verification}
\begin{enumerate}
\item How do you verify a claim? Please give an example. 
\item What are the principles do you follow in verifying a claim? For example, accuracy, non-partisanship, etc.
\item Do you follow any established guidelines for fact-checking? 
\item How do you ensure accuracy and fairness in verifying a claim? Please share an example.
\item How do you ensure transparency of sources? Do you use anonymous sources? Any example?
\item How do you ensure transparency of your methodology? For example, select, research, write, edit, publish and correct fact checks. 
\item Which experts do you contact for getting verification? What makes you think that they are expert in that area?
\item Do you select experts from your known circle? Have you considered getting opinion from experts who are not in your known circle? Did they provide information?
\item On average, how many people (of your organization) work on verifying a claim?
\end{enumerate}

\subsubsection{Challenges/Constraints}
\begin{enumerate}
\item What kind of challenges you face while doing your job?
\item Challenges while selecting what to check?
\item Challenges to check a claim?
\item Challenges in connecting to a source? Getting experts of a topic?
\item Challenges in dissemination?
\item Challenges in sustaining?
\item What technological challenges do you face? For example, unavailability of software, tools, etc.
\item What political challenges do you face? Please give examples.
\item Curtailed freedom of expression often makes a person to be self-censored. Have you faced such satiation while selecting and verifying a claim? 
\item What economic challenges do you face? Examples.
\item Any other challenges that you face?  
\end{enumerate}

\subsection{Interview Questionnaire for Journalists}
\subsubsection{General}
\begin{enumerate}
\item What is your duty with your news organization? What issues do you cover?
\item How long are you working in journalism?
\item Are you aware of fact-checking organizations? Either in Bangladesh or abroad?
\item If no, do you think politicians always say correct things?
\item If yes, what can journalists do in such instances?
\item Does your news organization have a separate fact-checking section?
\item If no, do you think your newspaper should have to combat the flood of fake information?
\item As a political/investigative journalist, whose and which claims do you verify?
\item How do you decide which sources you will talk to verify a claim? Please give examples.
\item How do you address selection bias in choosing your sources for verifying a claim? 
\item What steps do you take to provide enough details, sources and data to prove veracity of a claim? Please give an example. 
\item How do you assure that your verification process is nonpartisan and fair?
\end{enumerate}

\subsubsection{Awareness of fact-checking organizations in Bangladesh}  
\begin{enumerate}
\item Are you aware of any fact-checking organizations in Bangladesh? Do you follow their fact checks?
\item Have you cited their fact checks in your reports? If yes, please give examples; if no, why not?
\item Do you know any fact-checkers in Bangladesh? Have they contacted you for information such as getting access to sources?
\item Do you consider their fact checks reliable? Why or why not?
\item Fact-checkers verify claims to inform people. Do you consider their job as journalistic? Why or why not?
\item Sometimes fact-checkers verify claims published in mainstream media. Do you consider their work as friend or foe? Why or why not?
\item We constantly come across various claims, especially on social media. Do you think there is a need of independent fact-checking organizations beside news media to verify the claims?
\end{enumerate}


\end{document}